\documentclass[aps,twocolumn,nofootinbib]{revtex4}

\usepackage{graphicx}
\usepackage{amssymb, amsmath,mathtools}
\usepackage{bbm}
\usepackage{enumerate}
\usepackage{bbold}

\graphicspath{{./}{./plots/}}

\usepackage{xcolor}

\begin{document}

\title{p-wave superconductivity and the axi-planar phase of triple-point fermions}

\author{Subrata Mandal and Igor F. Herbut}

\affiliation{Department of Physics, Simon Fraser University, Burnaby, British Columbia, Canada V5A 1S6}

\begin{abstract}

We consider weak-coupling superconductivity in the inversion- and rotation-symmetric system of two pseudospin $s=1$ low-energy fermions of opposite chirality. General contact interactions can lead to Cooper instabilities towards d-wave or towards a novel $2\times 3$ p-wave matrix order parameter. We compute the Ginzburg-Landau free energy for the latter. Remarkably, in this case the Ginzburg-Landau free energy can be minimized exactly, with the resulting ordered state being analogous to the ``axi-planar" p-wave, which exhibits extra degeneracy, and generally breaks time reversal symmetry. Whereas a generic Ginzburg-Landau free energy for our order parameter has only $U(1) \times SO(2) \times SO(3)$ symmetry, at weak coupling we find it displaying the enlarged  $U(1)\times SO(3) \times U(1) \times SO(3)$ symmetry, broken down to $SO(2)\times SO(2)$ in the axi-planar ordered phase, and leading therefore to six Goldstone bosons. We show how the lattice, once restored, fixes the allowed values of the magnetization in the superconducting state by locking the spatial directions implicit in the order parameter to its high-symmetry axes.

\end{abstract}

\maketitle

Crystalline solids feature effective fermionic excitations outside the realm of possibilities of high-energy physics. Besides the ubiquitous Weyl fermions which have an effective ``spin" $s=1/2$ and describe crossing of two bands at a point, symmetry-protected quasiparticle  excitations with $s=1$ and $s=3/2$ are also possible. \cite{bradlyn} The latter Rarita-Schwinger-Weyl fermions have already been the subject of several investigations \cite{isobe, boettcher1, link1} that go beyond the non-interacting paradigm. Of great interest are also the triple-point fermions with $s=1$, which may arise when a symmetry-protected crossing of three bands occurs somewhere in the Brillouin zone. \cite{bradlyn, fulga, hu} The two linearly dispersing bands then carry the Chern number of $C = \pm 2$, whereas the flat band has $C=0$. To neutralize the topological charge there are then either (at least) two such points of triple-band crossings in the Brillouin zone with opposite helicities, or there are other topologically nontrivial bands that cross the Fermi level elsewhere in the Brillouin zone. The possible effects of electron-electron interactions on the triple-point fermions, to our knowledge, have not been studied yet. The recent detection of possible triple-point fermions in $CoSi$ \cite{rao} brings urgency to such investigations.

In a parallel development, Cooper pairing of multiband fermions has been systematically scrutinized recently, as it may naturally lead to multicomponent unconventional superconductivity with competing ground states, possible breaking of time reversal symmetry, or exotic quasiparticle spectra, exhibiting Bogoliubov-Fermi surfaces, for instance. \cite{agterberg, savary, boettcher2, lin, link, oh} Here we consider an inversion-symmetric system of two $s=1$ fermions of {\it opposite} chirality,  which represents the minimal, albeit non-relativistic extension of the Dirac Hamiltonian to higher spin. We show that a general contact attractive interaction may lead to superconducting matrix order parameter with p-wave symmetry,
which transforms under the Lie group $G = U(1)\times SO(2) \times SO(3) $. The $U(1)$ is the usual (particle number) gauge symmetry, the $SO(3)$ is the group of rotations, and the $SO(2)$ is the group of transformations between the two crossing points, or valleys, of opposite chirality. The crucial observation is that the p-wave order parameter is only a $2\times 3$ complex matrix because, as shown here, the opposite chirality of the two $s=1$ fermions cuts off the Cooper logarithm for three of the components of the usual $3\times 3$ matrix, which describes p-wave pairing in $^3$He, for example. \cite{vollhardt}

Interestingly, in spite of the large twelve-real-parameter space to which the order parameter  belongs, we find that the derived Ginzburg-Landau (GL) free energy for the  $2\times 3$ - component complex order parameter with the weak-coupling values of the coefficients of the five independent quartic terms \cite{mermin} can be minimized exactly; the minima break the $G$ symmetry completely, and exhibit an additional degeneracy, not implied by the symmetry $G$ of the GL theory, described by a single continuous parameter $\theta$. An equivalent p-wave configuration has appeared in the literature before as the so-called ``axi-planar" state \cite{barton},  but as far as we know it has not been previously found to be a minimum of any actual free energy. The ordered phase's extra degeneracy also manifests itself in the {\it six} Gaussian zero-energy modes around the axi-planar phase, instead of only five expected Goldstone bosons which would accompany the apparent symmetry breaking pattern. This hints at a possibly larger hidden symmetry of the GL theory, and we indeed find new variables that factorize the partition function into two terms, each describing a separate three-component complex order parameter coupled to its own chiral fermion. \cite{sim, mandal} At each of the two triple-band crossing points the superconducting condensate is maximally magnetized $j=1$ macroscopic quantum state of Cooper pairs, with average magnetization vector pointing along an arbitrary direction. The hidden symmetry breaking pattern is this way revealed as $U(1)\times SO(3) \times U(1)\times SO(3) \rightarrow SO(2) \times SO(2) $, which naturally accounts for the six observed massless modes by the standard Goldstone theorem.

Finally, we show that the degeneracy parameter $\theta$ is nothing but a half of the angle between the two symmetry axis in the $SO(2) \times SO(2)$ residual symmetry of the axi-planar state. This means that once a discrete lattice that harbors $s=1$ fermions is restored, $2\theta$ will be given by the angle between two equivalent symmetry-axes of the crystal. Reducing the rotational symmetry to the cubic symmetry, for example, leads to the preferred axes to be some of the cube's diagonals, and thus ultimately  the allowed values are $\theta=0, \pi/2, ArcTan[1 / \sqrt{2}], ArcCot[1 / \sqrt{2}]$. These translate into average {\it magnetization} of the p-wave superconducting state as maximal ($2$), minimal ($0$), or intermediate, $2\sqrt{2/3}$ or  $2 / \sqrt{3}$.

{\it Interacting triple-point fermions} -- Consider chiral $s=1$ fermions near two points in the Brillouin zone that exhibit triple-band crossing, with the low-energy Hamiltonian to the leading order in momentum,
\begin{equation}
H (\vec{p}) = \sigma_3 \otimes \vec{p} \cdot \vec{S},
\end{equation}
where the momentum is measured from the crossing points $\pm \vec{K}$, related to each other by inversion. The velocity has been set to unity.
The fermion field is a six-component Grassmann variable
\begin{equation}
\Psi(\omega, \vec{p}) = (a (\omega, \vec{K}+ \vec{p}), b (\omega, -\vec{K}+ \vec{p}))^T ,
\end{equation}
with $r = (r_1, r_2, r_3)^T$, ($r=a,b$), and $p \ll \Lambda$, with $\Lambda$ as the momentum cutoff, of the order of inverse of the lattice spacing. The matrices  $S_i$ ($i=1,2,3$) provide an $s=1$ (three-dimensional) representation of the $SO(3)$ Lie algebra, and the third Pauli matrix appearing in the left factor assures that the Chern numbers of the two degenerate eigenstates near the points $\pm \vec{K}$ cancel. The Hamiltonian is the direct sum of two $s=1$ Hamiltonians near the two crossing points, in exact analogy to the Dirac Hamiltonian as a sum of two $s=1/2$ (Weyl) Hamiltonians. It is inversion-symmetric, and commutes with the inversion operator $I=\sigma_1\otimes \mathbb{1}_3 $ accompanied by $\vec{p} \rightarrow - \vec{p}$. When (and only when) $\vec{K}=0$, it is also time-reversal-invariant, with time reversal operator $T= (\mathbb{1}_2 \otimes U)\cal{K}$ and $\vec{p} \rightarrow - \vec{p}$, where the unitary matrix $U$ is representation-dependent. Note however that since $s$ is integer $T^2 = U U^* = +  \mathbb{1}_6 $, and therefore $H (\vec{p})$ describes spinless (or spin-polarized) fermions. Alternatively, when $\vec{K}\neq 0$, $H (\vec{p})$ may be taken to describe regular half-integer-spin   fermions, but with broken time reversal symmetry.

We further assume a finite chemical potential $\mu$, and  $|\mu| \ll \Lambda$. The most important interactions are then those that are finite at the crossing points $\vec{p}=0$. We may parameterize them by contact interaction terms between the triple-band fermions which, for simplicity, are assumed to respect rotational invariance. The general interacting Lagrangian has then three such interaction terms \cite{herbut1}, and the form
\begin{widetext}
\begin{equation}
L = \Psi^\dagger(x) (\partial_\tau + H (-i\nabla) -\mu \mathbb{1}_6 ) \Psi (x)  - \sum_{r=s,p,d} V_r (\Psi^\dagger (x) M_r \Psi^* (x) )( \Psi^T (x)  M^\dagger _ r \Psi (x) ),
\end{equation}
\end{widetext}
where $x=(\vec{x}, \tau)$. Fermi-Dirac statistics requires the matrices $M_r$  to be antisymmetric, i. e.  $M_r ^T = - M_r $. In the adjoint representation $S_{i,jk} = - i  \epsilon_{ijk}$ they may be more conveniently written as $M_r = \Gamma_r (\sigma_2 \otimes \mathbb{1}_3 ) $, and recognize that the matrices $\Gamma_r$ provide then either the s-wave ($\Gamma_s = \mathbb{1}_2 \otimes \mathbb{1}_3 $), p-wave ($\Gamma_{p,ij}= \sigma_i \otimes S_j$), or d-wave ($\Gamma_{d, ij} = \mathbb{1}_2 \otimes T_{ij}$) pairing (with the five-component second-rank irreducible tensor $T_{ij} = \{S_i, S_j \} - (4/3) \delta_{ij} \mathbb{1}_3 $ \cite{boettcher2, mandal}). The repeated index summation convention is implied. We now take  $V_p > 0$, $V_s = V_d =0$, and consider the matrix order parameter
\begin{equation}
\Delta_{ij} (x)  = \langle \Psi^T (x) (\sigma_2 \sigma_i \otimes S_j) \Psi (x) \rangle,
\end{equation}
which would signal the formation of the p-wave superconductivity. We argue shortly that a weak coupling $V_s$ is irrelevant, whereas $V_d >0$ would be relevant, but lead to a familiar instability. \cite{boettcher2, mandal}

The complex matrix order parameter $\Delta_{ij}(x)$ is in general dependent on space and imaginary time, but for the purposes of the mean-field theory which we pursue here it suffices to take it to be uniform. The continuous symmetry of $L$ is evidently only $U(1) \times SO(2) \times SO(3) $, with the $U(1)$ being the particle number, $SO(2)$ inter-valley, and $SO(3)$ the rotation symmetry. Although the p-wave order parameter $\Delta_{ij}$ is a $3\times3$ matrix, the Lagrangian has the  symmetry smaller than the full rotational symmetry of $U(1) \times SO(3) \times SO(3)$, relevant to p-wave in $^3$He. The $2\times 3$ matrix $\Delta_{ij}$ ($i=1,2$, $j=1,2,3$) transforms as a two dimensional vector under the $SO(2)$ acting on the left index, whereas the remaining $\Delta_{3j} $ ($j=1,2,3$) is an $SO(2)$  scalar. Both objects transform as three-dimensional vectors under the $SO(3)$ rotations, acting on the right index.

{\it Ginzburg-Landau theory} -- Integrating out the fermions and expanding the result to two leading orders in powers of the uniform order parameter leads to the GL free energy $L_{GL} (\Delta) = L_2 (\Delta) + L_4 (\Delta)  + O(\Delta^6) $, where $L_2 (\Delta) $ is quadratic in the order parameter, and
\begin{equation}
L_2 (\Delta) = (\frac{\delta_{ij}}{V_p } + K_{ij} ) \Delta_{ik} ^*  \Delta_{jk}.
\end{equation}
Straightforward but somewhat lengthy calculation yields $K_{ij}=0$ for $i\neq j$, as required by the symmetry, and furthermore, at low temperatures $T$,
\begin{equation}
K_{11}= K_{22} = - \frac{2}{3} N_0 \ln (\frac{\Lambda}{T}),
\end{equation}
with $N_0 = \mu^2 / \pi^2$ as the density of states at the Fermi level. One observes the expected Cooper's logarithmic divergence as $T\rightarrow 0$. On the other hand, when $T\rightarrow 0$,
\begin{equation}
K_{33} = -\frac{2\Lambda^2}{3 \pi^2},
\end{equation}
and finite. Below the mean-field transition temperature $T_c = \Lambda e^{-3/(2 N_0 V)}$, $ \Delta_{ij} \neq 0$ only  for $i=1,2$ and $j=1,2,3$, whereas the remaining components $\Delta_{3i}=0$. In the limit of weak coupling, which is properly defined as $V_p \Lambda^2 \ll 1$,  the order parameter components $\Delta_{3i}$ remain massive at $T_c$, and even their fluctuations are suppressed at all temperatures.

The finiteness of $K_{33}$ follows also from noticing that the leading order effect of a finite $\Delta_{3i}$ would be only to shift the chemical potential, and therefore not open the usual gap {\it anywhere}. To see this, consider the corresponding Bogoliubov-de Genness Hamiltonian with the real \cite{remark} order parameter $ \Delta_{3i}$:
\begin{equation}
H_{BdG} = \sigma_3 \otimes (H(\vec{p} )-\mu \mathbb{1}_6) + \Delta_{3i} \sigma_2 \otimes \sigma_1 \otimes S_i.
\end{equation}
Although not obvious, it may be checked that
\begin{equation}
\det H_{BdG} = \mu^4 (p^2 - \mu^2- \Delta_{3i} \Delta_{3i})^4,
\end{equation}
and thus the entire Fermi surface is inflated from $p= \mu$ in the normal phase to $p= \sqrt{\mu^2 + \Delta_{3i} \Delta_{3i}}$ in the superconducting phase. The s-wave channel suffers from the same problem. The superconducting instabilities of the weakly interacting Lagrangian in Eq. (3) are therefore only the truncated $2 \times 3$ p-wave order considered here, and the d-wave studied before. \cite{boettcher2, mandal}

The precise form of the superconducting condensate below $T_c$ depends on the quartic terms in  $L_4 (\Delta) $:
\begin{equation}
L_4 (\Delta) = \sum_{i=1}^{5} b_i I_i (\Delta),
\end{equation}
where \cite{mermin}: $I_1 = |Tr \Delta \Delta^T|^2$, $I_2 = (Tr \Delta \Delta^\dagger)^2$, $I_3 = Tr \Delta \Delta^T (\Delta \Delta^T)^* $, $I_4 = Tr \Delta \Delta^\dagger \Delta \Delta^\dagger $, and $I_5 = Tr \Delta \Delta^\dagger (\Delta \Delta^\dagger)^* $. Although the order parameter is now only a $2 \times 3$ matrix, these five invariants under $G$ can be shown to still be linearly independent \cite{herbut1}. With arbitrary coefficients $b_i$  the $L_4 (\Delta) $ is the most general $G$-symmetric local quartic term. The standard one-loop diagram yields the {\it exact} relations between the quartic coefficients at all temperatures: $b_3 = -2 b_1$, $b_2 = b_4 = -b_5$, and $b_2 >0$ and $b_3 >0$. In addition, to the leading order in low $T_c$ one finds  also that $b_2 = b_3$, and all  $ b_i \sim N_0/T_c ^2$, as usual. \cite{zwerger} These turn out to be the same relations between the quartic coefficients as in the standard weak-coupling GL free energy for $^3$He.\cite{mermin}

{\it Mean-field solution and the axi-planar state} -- Writing the order parameter as $\Delta= \Delta_0 \Phi$, with $\Delta_0$ as the complex norm and $\Phi$ is a $2\times3$ normalized matrix satisfying  $Tr \Phi \Phi^\dagger =1$, the $L_{GL}$ is below $T_c$ minimized by
 \begin{equation}
 |\Delta_0|^2 = -\frac{a}{2 L_4(\Phi)},
 \end{equation}
 where $a= V_p ^{-1}  + K_{11}\sim (T-T_c)$. At this value of $\Delta_0$,
 \begin{equation}
 L_{GL} = -\frac{a^2}{4 L_4 (\Phi) },
 \end{equation}
 and therefore the GL free energy is the lowest at $\Phi$ that minimizes the expression
 \begin{equation}
 L_4 (\Phi) = b_2 + b_3 ( I_3 (\Phi) - \frac{1}{2} I_1(\Phi) ) + b_2 (I_4 (\Phi) - I_5 (\Phi) ).
 \end{equation}
Minimization of the last expression in terms of $3\times 3$ matrix would lead to the Balian-Werthamer state.\cite{vollhardt} We show next that in the present case $L_{GL} (\Phi)$ has a hidden symmetry and a rather different ordered phase.

 First, it is easy to see that both combinations of quartic invariants appearing in $L_4 (\Phi)$ are nonnegative. Since the matrix $\Phi \Phi^\dagger$ is $2\times 2$ and Hermitian, it can be expanded in terms of Pauli matrices: $\Phi \Phi^\dagger = c_\mu \sigma_\mu$, $\mu=0,1,2,3$, with real $c_\mu $. Then
 \begin{equation}
 I_4 (\Phi) - I_5 ( \Phi ) = Tr \Phi \Phi^\dagger (  \Phi \Phi^\dagger - (\Phi \Phi^\dagger)^T) = 4 c_2 ^2 \geq 0.
 \end{equation}
 Similarly, the non-Hermitian symmetric matrix $\Phi \Phi^T$ can be written as $\Phi \Phi^T = d_\mu \sigma_\mu$ with complex $d_\mu$, and $\mu\neq 2$. It readily follows that
\begin{equation}
I_3 (\Phi) - \frac{1}{2} I_1 (\Phi) = 2( |d_1 |^2  + |d_3|^2 ) \geq 0.
\end{equation}

Since both $b_2 >0$ and $b_3 >0$, the configurations for which $I_4 - I_5=I_3 - (I_1/2)=0 $, if existing, are the minima of the $L_{GL}$ at $T<T_c$.
To find such configurations, we  consider the two rows of the matrix $\Phi_{ij}$ as two (complex) vectors under the $SO(3)$: $\vec{\Phi}_1 = (\Phi_{11}, \Phi_{12},\Phi_{13})$, and  $\vec{\Phi}_2 = (\Phi_{21}, \Phi_{22},\Phi_{23})$. The normalization $Tr \Phi \Phi^\dagger =1$ then implies:
\begin{equation}
\vec{\Phi}_1 \cdot \vec{\Phi}_1 ^* +  \vec{\Phi}_2 \cdot \vec{\Phi}_2 ^* =1.
\end{equation}
The condition $I_3- (I_1/2) =0 $ demands that $\Phi \Phi^T $ is proportional to the unit matrix, which translates into
\begin{equation}
\vec{\Phi}_1 \cdot \vec{\Phi}_2 =0,
\end{equation}
\begin{equation}
\vec{\Phi}_1 \cdot \vec{\Phi}_1 = \vec{\Phi}_2 \cdot \vec{\Phi}_2.
\end{equation}
Likewise, the condition $I_4-I_5=0$ requires the matrix $\Phi \Phi^\dagger$ to be real, for which it suffices that its off-diagonal element
$\vec{\Phi}_1 \cdot \vec{\Phi}_2 ^*$ is real. However, a real symmetric matrix $\Phi \Phi^\dagger$ can always be rotated by an $SO(2)$ rotation into a diagonal form, in which a stronger but computationally more convenient condition applies:
\begin{equation}
\vec{\Phi}_1 \cdot \vec{\Phi}_2 ^* =0.
\end{equation}
A bit of algebra shows then that the most general solution of the Eqs. (16)-(19) (modulo a $G$-transformation, time reversal (complex conjugation), or shift of $\theta$) is given by
\begin{align}
 \Phi_\theta &= \frac{1}{\sqrt{2}  }  \begin{pmatrix} 0 & \sin{\theta} & 0  \\ 1 & 0 & i \cos \theta  \end{pmatrix},
 \end{align}
with an {\it arbitrary} $\theta$. This is analogous to the ``axi-planar" state, which may be understood as interpolating between the ``axial" ($\theta=0$) and ``planar" ($\theta=\pi/2$) phases.  \cite{barton}

Since the four invariants $I_i (\Phi) $, $i=1,3,4,5$, separately do depend on $\theta$, two configurations with different values of $\theta$ in general cannot be related by any symmetry transformation from $G$. The analysis of Gaussian fluctuations around $\Phi_\theta$ confirms this: one finds {\it six} independent massless modes, whereas only five are implied by the symmetry breaking pattern of $\Phi_\theta$, which for a general $\theta$ breaks all five generators of $G$.

{\it Separation variables}--The hidden symmetry of $L_{GL}$ is revealed by the transformation from $\vec{\Phi}_{1,2}$ into $\vec{\Phi}_{\pm}$, defined as
$\vec{\Phi}_s = \vec{\Phi}_{2} +i s \vec{\Phi}_{1}$, $ s=\pm $. In terms of these variables GL free energy becomes $L_{GL} = L_+ + L_-$, where
\begin{equation}
L_s = \frac{a|\Delta_0|^2}{2}+ \frac{b_3 |\Delta_0|^4}{ 2} ( 1 + \frac{1}{2} |\vec{\Phi}_{s} \cdot\vec{\Phi}_{s}|^2 ),
\end{equation}
and we imposed the normalization $\vec{\Phi}_{s}^* \cdot\vec{\Phi}_{s} =1$ (with no summation over $s$ implied hereafter), and assumed $b_2 = b_3$ in Eq. (11), as appropriate at low temperatures. It is evident that each complex $SO(3)$ vector order parameter $\vec{\Phi}_{s}$ can be gauge-transformed and rotated independently of the other, so the hidden symmetry of $L_{GL}$ is in fact $U(1)_+ \times SO(3)_+ \times U(1)_- \times SO(3)_- $. Each condensate $\vec{\Phi}_{s}$ represents a spin-one macroscopic quantum state $|\Phi_s\rangle$, and is found to couple only to the triple-point fermion of chirality $s$. Each $L_s$ is minimized by the state for which $\vec{\Phi}_{s} \cdot\vec{\Phi}_{s}=0$. Since in the adjoint representation,
\begin{equation}
|\vec{\Phi}_{s} \cdot\vec{\Phi}_{s}|^2 = 1 - ( \Phi_{s,n}^* S_{i,nm}  \Phi_{s,m}  ) ( \Phi_{s,k}^* S_{i,kj}  \Phi_{s,j}  ),
\end{equation}
each $\vec{\Phi}_s$ maximizes its possible average magnetization $\vec{m}_s = \langle \Phi_s|\vec{S} | \Phi_s \rangle$, i. e. it represents the $m=1$ normalized state along some axis $\hat{n}_s $. Taking $\hat{n}_s ^2 =1$, the eigenvalue equation $(\hat{n}_s \cdot \vec{S})_{ij} \Phi_{s,j} = \Phi_{s,i} $ becomes
\begin{equation}
\hat{n}_s \times \vec{\Phi}_s = i \vec{\Phi}_s.
\end{equation}
Choosing $\hat{n}_s = (0, \cos \theta, (-s)\sin \theta)$ in the $x=0$ plane, for example, one finds $\vec{\Phi}_s = (1, (i s) \sin\theta, i\cos\theta)/\sqrt{2}$, which yields precisely the matrix in Eq. (20). The parameter $\theta$ is thus nothing but a half of the angle between the two independent  axes $\hat{n}_+$ and $\hat{n}_-$. The average magnetization in the ordered state is $\hat{n}_+ + \hat{n}_- $, and therefore the parameter $\theta$ directly determines its magnitude, $ 2 |\cos \theta|$.

{\it Lattice symmetry} -- Whereas the directions of the axes $\hat{n}_{\pm}$ are arbitrary in the continuum limit, it is no longer so once the lattice is restored. The most general quartic term in each $L_s$ that respects cubic symmetry, for example, is
\begin{equation}
L_{s4} \propto |\Delta_0|^4 ( q_1+ q_2 |\vec{\Phi}_{s} \cdot\vec{\Phi}_{s}|^2  + q_3 \sum_{i=1} ^3 | \Phi_{s,i} |^4).
\end{equation}
For $q_2 > -q_3/2$, $\vec{\Phi}_s$ with maximal average magnetization is still the minimum of this GL free energy, but the axis of magnetization depends on the sign of $q_3$: for $q_3 >0$ it is $\hat{n}_s = (\pm 1,\pm 1,\pm 1)/\sqrt{3}$, i. e. one of the cube's diagonals, whereas for $q_3 <0$ it is  $\hat{n}_s = (0,0,1)$ and permutations thereof, i. e. one of the $C_4$-axes of symmetry. Reducing the symmetry to cubic by replacing $p_i \rightarrow \sin p_i$  ($i=1,2,3$) in Eq. (1), and then rederiving the GL free energy yields $q_3 >0$ for $s=\pm$. The two axes $s=\pm$,  besides being identical or opposite, can thus also be along two different diagonals of the cube. This leads to the discrete values of $\theta$ and the concomitant average magnetization quoted in the introduction.

{\it Final remarks} -- The axi-planar phase breaks time reversal symmetry, unless $\theta=\pi/2$. However, even in this case, time reversal is broken at each valley $\pm \vec{K}$ individually, but in the opposite ways. Since the valleys are decoupled by the axi-planar order parameter, the quasiparticle spectrum features mini Bogoliubov-Fermi surfaces \cite{sim}, as expected from general considerations.\cite{link}

Another consequence of the decoupling of the valleys is that the partition function at each valley is analogous to the one for frustrated magnets, and the finite-temperature phase transition is most likely weakly first-order \cite{boettcher3}. One also finds that the weak-coupling relations between quartic coefficients, namely $b_3 = -2 b_1$ and  $b_2 = b_4 = -b_5$, remain preserved under the renormalization group flow, \cite{jones, unpub} exactly as mandated by the enlarged  symmetry that protects them.

{\it Acknowledgements} -- The authors are grateful to Igor Boettcher and Julia Link for many useful discussions. This work was supported by the NSERC of Canada.


\begin{thebibliography}{99}

\bibitem{bradlyn} B. Bradlyn, J. Cano, Z. Wang, M. G. Vergniory, C. Felser, R. J. Cava, and B. A. Bernevig, Science {\bf 353}, aaf5037 (2016).



\bibitem{isobe} H. Isobe and L. Fu, Phys. Rev. B {\bf 93}, 241113(R) (2016).
\bibitem{boettcher1} I. Boettcher, Phys. Rev. Lett. {\bf 124}, 127602 (2020); Phys. Rev. B {\bf 102}, 155104 (2020).
\bibitem{link1} J. M. Link, I. Boettcher, I. F. Herbut, Phys. Rev. B {\bf 101}, 184503 (2020).

\bibitem{fulga} I. C. Fulga and A. Stern, Phys. Rev. B {\bf 95}, 241116(R) (2017).
\bibitem{hu} H. Hu, J. Hou, F. Zhang, and C. Zhang, Phys. Rev. Lett. {\bf 120}, 240401 (2018).

\bibitem{rao} Z. Rao, et al., Nature  {\bf 567}, 496 (2019).

\bibitem{agterberg} D. F. Agterberg, P. M. R.  Brydon, and C. Timm, Phys. Rev. Lett. {\bf 118}, 127001 (2017); P. M. R. Brydon, D. F. Agterberg, H. Menke, and C. Timm, Phys. Rev. B {\bf 98}, 224509 (2018).
\bibitem{savary} L. Savary, J. Ruhman, J. W. F. Venderbos, L. Fu, and P. A. Lee, Phys. Rev. B {\bf 96}, 214514 (2017); J. W. F. Venderbos, L. Savary, J. Ruhman, P. A. Lee, and L. Fu, Phys. Rev. X {\bf 8}, 011029 (2018).
\bibitem{boettcher2} I. Boettcher and I. F. Herbut, Phys. Rev. Lett. {\bf 120}, 057002 (2018); Phys. Rev. B {\bf 93}, 205138 (2016).
\bibitem{lin} Y.-P. Lin and R. M. Nandkishore, Phys. Rev. B {\bf 97}, 134521 (2018).
\bibitem{link} J. M. Link and I. F. Herbut, Phys. Rev. Lett. {\bf 125}, 237004 (2020); I. F. Herbut and J. M. Link, Phys. Rev. B {\bf 103}, 144517 (2021).
\bibitem{oh} H. Oh and E.-G. Moon, Phys. Rev. B {\bf 102}, 020501(R) (2020).

\bibitem{vollhardt} D. Vollhardt and P. Wolfle, {\sl The Superfluid Phases of Helium 3} (Taylor and Francis, London, 1990).
\bibitem{mermin} N. D. Mermin and D. Stare, Phys. Rev. Lett. {\bf 30}, 1135 (1973).
\bibitem{barton} G. Barton and M. Moore, J. of Phys. C: Sol. St. Phys., {\bf 7}, 4220 (1974); N. D. Mermin and D. Stare, 1974 Cornell preprint.
\bibitem{sim} G. Sim, M. J. Park, and S. Lee, arXiv:1909.04015.
\bibitem{mandal} S. Mandal, J. Link, and I. F. Herbut, arXiv:2105.07568.


\bibitem{herbut1} I. F. Herbut, Phys. Rev. D {\bf 100}, 116015 (2019).

\bibitem{remark} Since the GL free energy to the quadratic order depends only on $\Delta_{3i} ^* \Delta_{3i}$, to detect Cooper instability it suffices to assume real $\Delta_{3i}$.

\bibitem{zwerger} S. Stintzing and W. Zwerger, Phys. Rev. B {\bf 56}, 9004 (1997); I. F. Herbut, Phys. Rev. Lett. {\bf 85}, 1532 (2000).

\bibitem{boettcher3} I. Boettcher and I. F. Herbut, Phys. Rev. B {\bf 97}, 064504 (2018), and references therein.
\bibitem{jones} D. R. T. Jones, A. Love, and M. A. Moore, J. Phys. C {\bf 10}, 1159 (1977).
\bibitem{unpub} S. Mandal and I. F. Herbut, unpublished.


\end{thebibliography}
\end{document}